\documentclass[times]{elsarticle}
\journal{Journal of \LaTeX\ Templates}

\usepackage{lineno,hyperref}

\usepackage{amsmath}
\usepackage{setspace}
\usepackage{array}
\usepackage{booktabs}
\usepackage{graphicx}
\usepackage[export]{adjustbox} 

\usepackage{subcaption}
\usepackage[rgb]{xcolor}
\usepackage{tikz}
\usepackage{pgfplots}

\usepackage{verbatim}
\usepackage{enumerate}

\usepackage{siunitx}
\usepackage{textcomp}
\usepackage{chemfig}

\newcommand{\matchsupheight}{_{\vphantom{0}}}

\usepackage[utf8]{inputenc}

\usepackage{float}

\mathchardef\mhyphen="2D

\begin{document}
\fontdimen14\textfont2=5pt

\newcommand{\hwkonnov}{\raisebox{2pt}{\tikz{\draw[newOrange,solid,line width=0.9pt](0,0) -- (5mm,0);}}}
\newcommand{\hwUCSD}{\raisebox{2pt}{\tikz{\draw[newBlue,solid,line width=1.2pt](0,0) -- (5mm,0);}}}
\newcommand{\hwUSC}{\raisebox{2pt}{\tikz{\draw[newRed,solid,line width=1.2pt](0,0) -- (5mm,0);}}}
\newcommand{\hwli}{\raisebox{2pt}{\tikz{\draw[newGreen,solid,line width=1.2pt](0,0) -- (5mm,0);}}}
\newcommand{\hwboivin}{\raisebox{2pt}{\tikz{\draw[newGray,solid,line width=1.2pt](0,0) -- (5mm,0);}}}

\begin{frontmatter}

\title{Anchoring of Turbulent Premixed Hydrogen/Air Flames at Externally Heated Walls}

\tnotetext[mytitlenote]{Fully documented templates are available in the elsarticle package on \href{http://www.ctan.org/tex-archive/macros/latex/contrib/elsarticle}{CTAN}.}

\author{Sebastian Klukas}
\author{Marcus Giglmaier}
\author{Nikolaus A. Adams}
\address{Technical University of Munich}

\author{Moritz Sieber}
\author{Sebastian Schimek}
\author{Christian O. Paschereit}
\address{Technical University of Berlin}

\begin{abstract}
A joint experimental and numerical investigation of turbulent flame front anchoring at externally heated walls is presented. The phenomenon is examined for lean hydrogen/air mixtures in a novel burner design, which comprises a cylindrical burning chamber with an integrated wall heating assembly converging into a quartz glass pipe. The transparent part allows for optical \chemfig{OH^{*}} chemiluminescence measurements serving as a basis for numerical validation. Besides being an often required prerequisite for continuous combustion devices, the investigated flame anchoring mechanism is intended to stabilize secondary flames in multi-fuel  combustion applications. \newline
For a comprehensive numerical evaluation the effect of heat loss on different hydrogen/air chemical reaction mechanisms is reviewed in a preparatory one-dimensional flame study. The subsequent numerical investigation is conducted by solving the compressible Reynolds-averaged Navier-Stokes (RANS) equations in axisymmetric domains. It focuses on the application of the eddy dissipation concept (EDC) as a tur{\-}bulence-chemistry interaction model in the realm of wall anchoring turbulent flames. The influence of different two-equation turbulence models and EDC modeling constants are discussed. Since wall heat transfer is responsible for ignition as well as quenching of the flame front, a special focus is put on boundary layer resolving near-wall treatment. A qualitative comparison between simulations and experiment is performed for multiple operating conditions. These are selected to display the influence of equivalence ratio, bulk Reynolds number and unburnt mixture temperature. \newline 
While the choice of RANS-based turbulence model has a distinguishable impact, EDC modeling coefficients exhibit a more significant influence on flame shape and length. It is only surpassed by the impact of correct diffusion treatment on reacting lean hydrogen/air mixtures. Therefore, full multicomponent diffusion treatment using the Maxwell-Stefan equation is applied. Adapted EDC modeling constant are employed for all operating conditions and only display pronounced discrepancies for small unburnt mixture temperatures. 

\end{abstract}


\end{frontmatter}

\section{Introduction}
\label{sec:introduction}
Controlled flame anchoring is a regular prerequisite for continuous combustion devices. Exemplary applications range from  micro-combustors over jet engines to scramjets \cite{wan2018} \cite{fureby2000} \cite{gruber2001}.
In general, flame anchoring may be realized by three different means of flow control: recirculation, supersonic shocks or heat transfer. Recirculation-based flame holding is either be induced by wall cavities \cite{benYakar2001} \cite{wan2015} or bluff bodies \cite{wan2018} \cite{kedia2014}, which both come in various shapes and forms. Shock-induced combustion applications utilize the enhanced thermodynamic state after a shock wave to continuously ignite a mixture \cite{brouillette2017} \cite{goertz2011} \cite{giglmaier2013}. Additionally, heat abduction, for example in form of porous media \cite{dunnmon2017}, or heat introduction by external means may lead to steady flame anchoring. While the latter concept has been comprehensively studied for laminar flames and micro-combustion \cite{sanchezSanz2014} \cite{maruta2005} \cite{bioche2019}, it is relatively unexplored for turbulent flames in the macroscopic scale. This study aims to investigate the phenomenon thoroughly by numerical and experimental means. \newline
Flame stabilization by supplementary heat introduction is categorized as excess enthalpy combustion \cite{kotani1982}. It describes the behavior of a flamelet in a continuously perfused duct or pipe and especially its thermal interplay with the adjacent walls. In this configuration excess heat of the burnt gas is conducted through the wall back to the unburnt gas, thereby heating the unburnt mixture and extending its flamability limit. The detailed behavior depends on the heat transfer coefficient between fluid and solid, heat conduction capability of the solid as well as its thermal heat capacity. Without any external heating capability, the flame front tends to advance back and forth as well as extinguish and re-ignite periodically \cite{maruta2005}. To remedy this issue, additional external heat is introduced locally into a specified wall segment. This leads to a less sensitive flame-wall interaction and allows for spatial control of the flame front position. The general effect has been studied with a particular focus on laminar flames \cite{bioche} \cite{maruta2005}.\newline
The study's objective is to examine the presented flame anchoring mechanism both experimentally and numerically. This requires special attention on turbulent flame fronts and their inherent interplay between turbulence, reaction pace and wall quenching. To create a well understood test case a novel burner design is developed. It features an elevated turbulence degree inferring highly non-uniform local flame speeds which makes quasi-steady flame anchoring challenging. This obstacle can be overcome by the employed flame anchoring mechanism. In the numerical analysis the interaction of turbulence and chemistry is prioritized by evaluating the species' reaction terms in a RANS-context by the well-established eddy dissipation concept. \newline
One of the benefits of excess enthalpy flame holders is that very lean stable burning conditions are attainable and hence, their potential for ultra-lean combustion is evaluated. Furthermore, the low equivalence ratio reduces flame speed and allows for moderate flow velocities as well as bulk Reynolds numbers. In addition, it leads to weakened \chemfig{NO_{x}} production and flame temperatures benefiting corrosion behavior as well as environmental concerns. Hydrogen fuel is employed and combined with air as an oxidizer. While being one of the most interesting fuels for future clean energy solutions, hydrogen combustion comprises very high flame speeds. Thus, the experimental burner features elevated Reynolds numbers besides operating under ultra-lean air-fuel ratios. \newline
After presentation of the employed burner setup and its operating conditions, the paper is structured as follows. First, a review of the applied numerical method and especially the EDC is exercised. Secondly, different kinetic mechanisms for hydrogen/air combustion are examined and compared with respect to their ability to portray heat losses as induced by adjacent walls. Thirdly, the influence of different turbulence models and EDC modeling parameters on flame shape and simulation results in general are discussed. Finally, varying operating conditions displaying the impact of equivalence ratio, bulk Reynolds number as well as unburnt mixture temperature are reviewed and utilized for experimental validation of the numerical setup. 

\section{Experimental setup}
To investigate anchoring of turbulent flames at preheated walls, a novel burner design is implemented and tested thoroughly. The device is constructed to operate under a wide range of equivalence ratios, inlet mass flow rates and unburnt mixture temperatures. The detailed experimental setup, employed measurement techniques and operating conditions are described in this section. \newline 

\subsection{Burner design and instrumentation}
\begin{figure}[htb]
	\centering
	\hspace*{-3.5cm}
	\includegraphics[width=1.6\textwidth]{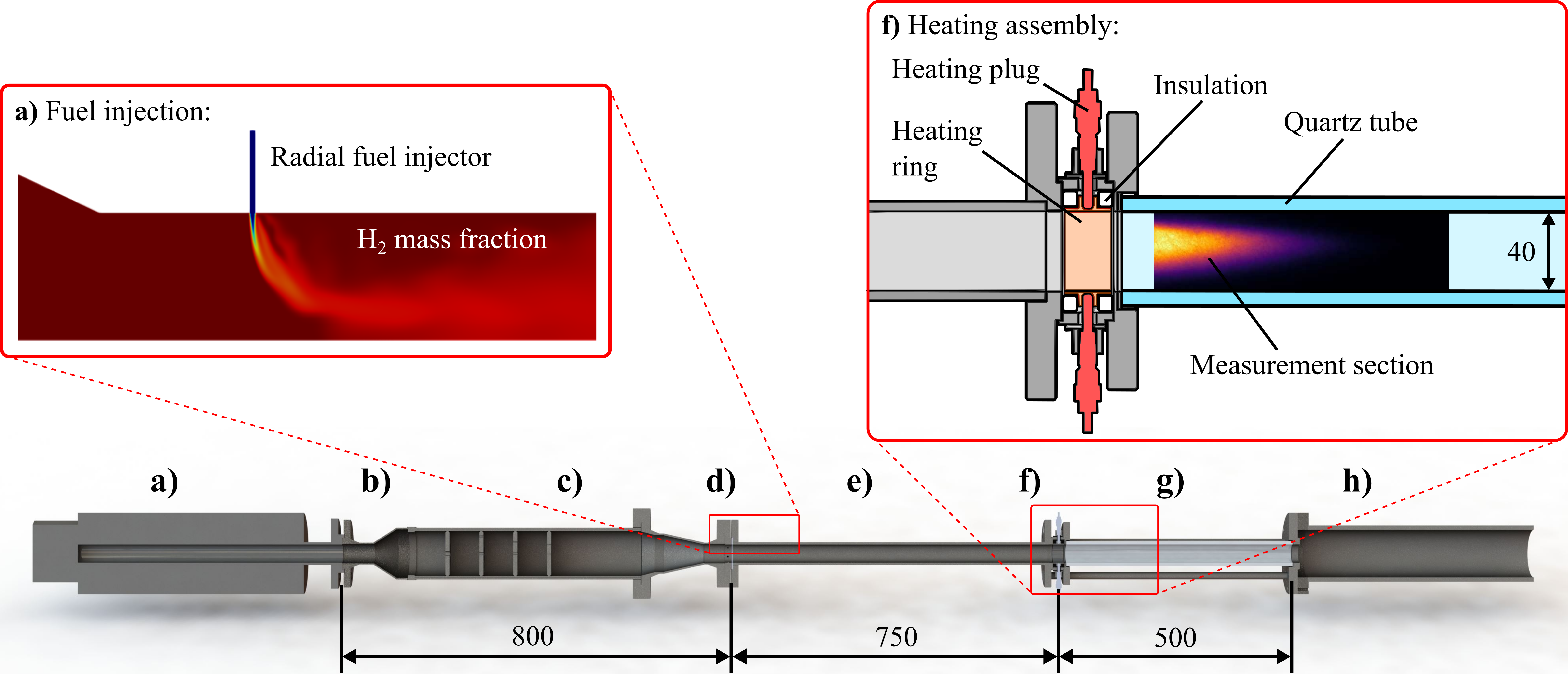}
	\caption{Schematic of the experimental burner setup. Details of the heating assembly and fuel injection are visualized in separate magnified sections.}
	\label{fig:schematic}
\end{figure}

The burner design features the following sections and their respective function as they are displayed in figure \ref{fig:schematic}:
\begin{enumerate}[a)]
	\itemsep0em 
	\item An electric preheater provides hot air at a pressure of $\SI{111}{\bar}$ and adjustable temperature between $\SI{200}{\celsius}$ and $\SI{400}{\celsius}$.
	\item The flow is then decelerated via a divergent nozzle before a 
	\item set of narrow narrow honey combs grids is applied to diminish turbulent fluctuations.
	\item After the straightened flow is reaccelerated by a converging nozzle, hydrogen fuel is inserted radially through six injection ports, equally spaced in circumferential direction.
	\item Downstream, a long steel tube ensures that fuel and oxidizer have enough time to fully mix, which justifies the assumption of a premixed flame. Successful mixing has been confirmed by non-reacting three-dimensional CFD simulations.
	\item The heating assembly, which functions as a flame holder, is located at the very end of the steel tubing and provides the necessary means for external heating of a separate wall section, which eventually leads to ignition. 
	\item Subsequently, a quartz glass tube makes the flame front visually accessible by the measuring equipment. 
	\item Finally, a water cooled exhaust pipe is utilized to chill down the hot reaction products before they are released into the environment.
\end{enumerate}
The heating assembly comprises four electric heating plugs connected to a heating ring as well as insulation bands (see top right of fig. \ref{fig:schematic}). Therefore, the heating ring partially replaces the chamber wall while insulation inhibits heat conduction to adjacent parts. Equal spacing of the heating plugs ensures an uniform temperature distribution within the ring and thus symmetrical flame positioning. Hence, a distinctive locally heated wall is devised. The design is optimized to avoid any wall gaps in order to diminish disturbance of the turbulent boundary layer. 
\newline
The total axial dimension of the burner, excluding the preheater and exhaust, is $\SI{2.05}{\meter}$ and the respective wall thicknesses of the steel tube and glass pipe are $\SI{5}{\milli\meter}$ and $\SI{7.5}{\milli\meter}$. The metal part upstream of the heating assembly is thermally isolated with stone wool. Temperature measurements within the burner confirm that no significant wall heat loss takes place. The installed heating ring extends for $\SI{24}{\milli\meter}$ and is placed $\SI{20}{\milli\meter}$ in front of the glass tube. \newline
The measurement setup provides the means for a qualitative comparison between numerics and experiment. For this purpose relevant boundary conditions as well as the flame's reaction zone are quantified and evaluated. Air and hydrogen inlet mass flow rates are registered by a set of integrated mass flow controllers. Temperature measurements of the mixture are obtained by thermocouples placed along the burner's axis. The heating ring is equipped with additional thermocouples to determine the local wall temperature. Equal temperature distribution inside the ring is confirmed visually. To assess general flame shape and quantify the position of the reaction front, a digital camera is utilized. By appliance of an optical filter transmitting a wavelength of $\SI{308}{\nano\meter}$ the hydroxyl radical (\chemfig{OH^{*}}) distribution becomes discernable. These chemiluminescence measurements are averaged over $\SI{200}{\micro\second}$ to give an accurate representation of flame shape and reaction zone. 

\subsection{Operating conditions}
\label{subsec:operatingConditions}
The influence of three major reactive flow characteristics and their effect on burner stability and overall flame shape are investigated. Those are: equivalence ratio, inlet mass flow rate and unburnt mixture temperature. The inlet mass flow rate primarily influences the bulk Reynolds number. The experimental setup described above allows precise tuning of these variables. To illustrate their respective influence, five operating points are considered as displayed in table \ref{tab:operatingConditions}. 
\begin{table}[H]
	\caption{Featured operating conditions determined by the mixture's total mass flow rate $\mathrm{\dot{m}_{total}}$, its temperature prior ignition $\mathrm{T_{u}}$, equivalence ratio $\mathrm{\Phi}$, bulk Reynolds number $\mathrm{Re}$, turbulent Reynolds number $\mathrm{Re_t}$ as well as turbulent Damk{\"o}hler number $\mathrm{Da_t}$. }
	\centering
	\def\arraystretch{1.3}
	\begin{spacing}{1.1}\centering
		\begin{tabular}{@{} c  >{\centering\arraybackslash}m{1.cm} >{\centering\arraybackslash}m{0.8cm}  >{\centering\arraybackslash}m{0.7cm} >{\centering\arraybackslash}m{1.cm} >{\centering\arraybackslash}m{1.cm} c@{}}
			\toprule
			& $\mathrm{\dot{m}_{total}}$ & $\mathrm{T_{u}}$ & $\mathrm{\Phi}$ & $\mathrm{Re}$ & $\mathrm{Re_t}$  & $\mathrm{Da_t}$ \\
			Case & $[\si{\kg\per\hour}]$ & $[\si{\celsius}]$ & $[\mhyphen]$ & $[\mhyphen]$ & $[\mhyphen]$ & $[\mhyphen]$ \\
			\midrule 
			1 & 80.47 & 400 & 0.2 & 21463 & 69.1 & 1752 \\
			2 & 80.67 & 400 & 0.3 & 21555 & 69.4 & 1745 \\
			3 & 60.5  & 400 & 0.3 & 17137 & 56.7 & 2228 \\
			4 & 60.35 & 400 & 0.2 & 16096 & 53.7 & 2254 \\
			5 & 80.47 & 300 & 0.3 & 23971 & 76.1 & 2710 \\
			6 & 80.47 & 200 & 0.3 & 27338 & 85.4 & 4610 \\
			\bottomrule
		\end{tabular}
	\end{spacing}
    \label{tab:operatingConditions}
\end{table}
Turbulent premixed combustion is categorized by its flame front behavior, specifically the reaction's interaction with turbulence \cite{peters1999} \cite{poinsot2005}. Based on turbulent Reynolds number $\mathrm{Re_t}$ and turbulent Damk{\"o}hler number $\mathrm{Da_t}$ various premixed combustion regimes are characterized. For the defined operating conditions the burner's combustion behavior is placed in the quasi-steady corrugated flamelets regime ($\mathrm{Da_t \gg 1}$ and $\mathrm{Re_t \gg 1}$). In this regime turbulence-chemistry interaction mandates to be taken into account for all numerical calculations. 

\section{Numerical setup and theory}
To numerically investigate the presented flame-anchoring mechanism, the fully compressible Reynolds-averaged Navier-Stokes (RANS) equations are solved by the commercial CFD code Ansys Fluent in version $\mathrm{2019\;R3}$ \cite{ansys}. One of the challenges inherent to turbulent time-averaged combustion calculations is the determination of each species' reaction rate. Since turbulence driven mixing processes affect those rates drastically, the eddy dissipation concept is utilized as a modeling approach to evaluate the species' reaction source. Besides a description of the general numerical setup, the combustion model is briefly reviewed in the following. \newline

\subsection{Numerical setup}
\begin{figure}[H]
	\centering
	\includegraphics[width=0.8\textwidth]{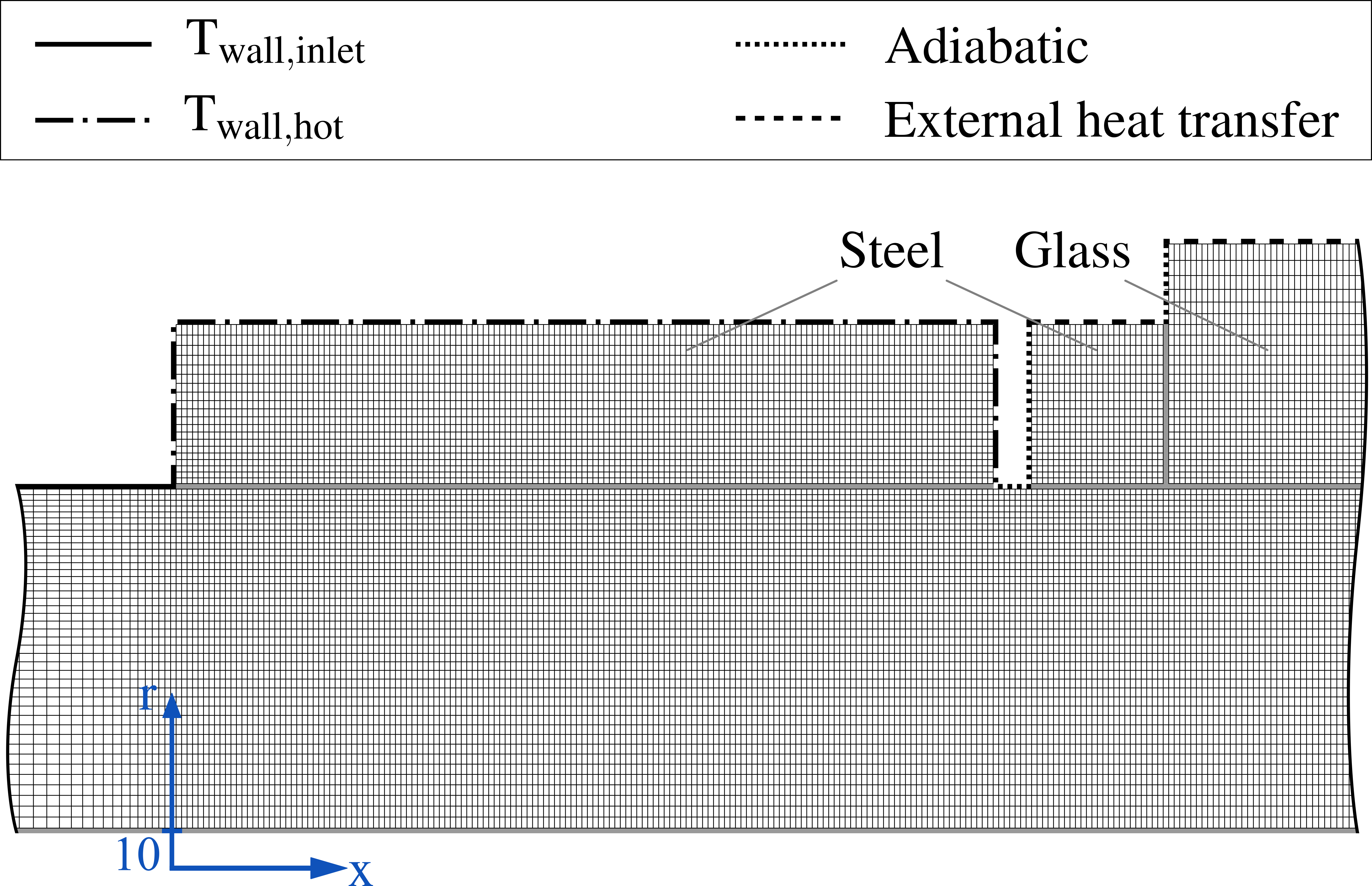}
	\caption{Grid design and implemented thermal boundary conditions. In radial direction only a segment with $\mathrm{r \geq 10\; \si{\milli\meter}}$ is displayed. For enhanced clarity the cell density is reduced by a factor of four.}
	\label{fig:grid}
\end{figure}

To achieve a steady-state solution, a fully-coupled implicit solution procedure, utilizing an algebraic multigrid pressure-based finite volume solver, is applied. The pseudo-transient implicit under-relaxation formulation is employed for a stable and efficient convergence behavior. Spatial discretization is performed by the quadratic interpolation upwind (QUICK) scheme ensuring second order accuracy \cite{leonhard1979}, whereas the pressure equation is discretized by the PRESTO! scheme. The closure problem of the time-averaged Navier-Stokes equations is addressed by established $\mathrm{\omega\mhyphen}\text{based}$ eddy viscosity as well as Reynolds stress turbulence models. A detailed description is omitted at this point but has been addressed at length by \cite{pope2000}. \newline
Although buoyancy may impact the shape of wall anchoring flames \cite{bioche20192}, gravity effects are assumed negligible. This deduces the assumption of axisymmetric flow conditions, which is confirmed by \chemfig{OH^{*}} chemiluminescence measurements during the experimental test campaign. Since heat conduction in the burner's walls plays a major role in the flame anchoring mechanism \cite{bioche2019}, they are considered in the simulations by solving Fourier's heat conduction equation. \newline
Thermochemical fluid properties are evaluated by the CHEMKIN software libraries \cite{chemkin}. Therefore, heat capacity is assessed by a fourth order polynomial fit while heat conductivity, viscosity and molecular as well as thermal diffusion coefficients are derived from kinetic theory. Mixture averaging is performed for all transport properties except diffusion. Bruno et al. \cite{bruno2015} concluded that multicomponent diffusion plays an essential factor in high Reynolds number reactive flows. Especially for hydrogen/air mixtures and their inherently high diffusion coefficient of hydrogen, the overall flow field may be crucially impacted by differential diffusion \cite{chen2007}. Therefore, binary diffusion coefficients are calculated by the Dixon-Lewis method and adapted into the flow by the Maxwell-Stefan diffusion equation. 
While the fluid's properties are highly dependent on local flow state, solid material attributes are assumed to be constant and are summarized in table \ref{tab:solidProperties}. 
\addtolength{\tabcolsep}{-3pt}    
\begin{table}[H]
	\caption{Solid material and radiation properties.}
	\centering
	\def\arraystretch{1.3}
	\begin{spacing}{1.1}\centering
		\begin{tabular}{@{} l  c c c c c   @{}}
			\toprule
			& $\mathrm{\rho} $ & $\mathrm{c_p}$ & $\mathrm{\lambda}$ & $\mathrm{\epsilon}$ & $\mathrm{f_d}$   \\
			& $[\si{\kg\per\meter\cubic}]$ & $[\si{\joule\per\kg\per\kelvin}]$ & $[\si{\watt\per\meter\per\kelvin}]$ & $[\mhyphen]$ & $[\mhyphen]$  \\
			\midrule 
			Stainless steel & 7900 & 750 & 25   & 0.8 & 0.5 \\
			Quartz glass    & 2200 & 740 & 1.38 & 0.8 & 0.1 \\
			\bottomrule
		\end{tabular}
	\end{spacing}
	\label{tab:solidProperties}\end{table}
\addtolength{\tabcolsep}{+3pt}    
Radiation phenomena play a crucial role in the analysis of high temperature reacting flows. Therefore, the discrete ordinates radiation model is applied in the presented configuration. The amount of solved transport equations for radiation intensity depends on the angular discretization, which in our case equals 24 control angles. The radiation energy transfer equations are solved in a sequential manner. All surfaces are considered to be gray walls. Special attention is required for the semi-transparent quartz-glass which is the only non-opaque surface in the domain. Therefore, radiation has to be treated as a volumetric and not a pure surface phenomenon. In general, radiation is considered to be partially diffusive and partially specular. The applied internal emissivities $\epsilon$ and diffusive fractions $f_d$ are listed in table \ref{tab:solidProperties}. \newline
The computational domain extends from the radial hydrogen injection until the end of the glass tube. But, radial fuel injection is not reflected in the simulation. Instead, the mixture is assumed to be perfectly premixed at the domain inlet. This assumption has been verified by preceding cold gas mixing simulations. By considering a long inlet in front of the flame holder, it is ensured that profiles of velocity, turbulent kinetic energy and its dissipation rate are fully developed prior to ignition. Since the fluid's temperature upstream of the flame holder remains constant, corresponding Dirichlet boundary conditions are applied (fig. \ref{fig:grid}). \newline
To reflect the wall anchoring mechanism correctly, it is crucial to match the behavior of the heating ring assembly in the numerical setup. As a necessary simplification, solely the heating ring is considered as an adjacent wall. In that way the physical phenomenon is still well represented, without the need to include the heating assembly's complicated geometry (fig. \ref{fig:grid}). Isolation bands, which inhibits heat transfer to neighboring walls, are approximated by an adiabatic boundary condition and missing wall connection. An overview of applied boundary conditions is visualized in figure \ref{fig:grid}.  The axial point of origin ($\mathrm{x = 0}$) is placed right at the beginning of the heated wall segment to simplify comparison between experimental and numerical data. \newline
The block structured hexahedral mesh is designed around two objectives: First and foremost the condition $y^+ \approx 1$ has to be met to correctly resolve the turbulent wall boundary layer for all operating conditions. Secondly, the mesh resolution has to be fine enough to reflect the thin reaction zone and its quickly recombining intermediate species. To satisfy those requirements, wall adjacent cells feature a radial extend of $40\; \si{\micro\meter}$ and unity aspect ratio. While the aspect ratio is kept constant along the externally heated section, cells grow exponentially in all axial and radial directions. Finally, the grid design comprises $1.49\cdot10^6$ cells in total.

\subsection{Eddy dissipation concept}
A critical challenge in turbulent RANS combustion simulations poses the correct calculation of each species' mean reaction rate. Since  an evaluation utilizing an Arrhenius approach combined with mean flow quantities leads to significant deviations, a modeling strategy to incorporate turbulence-chemistry interaction must be applied \cite{poinsot2005}. One appropriate model applicable to turbulent premixed flames is the eddy dissipation concept (EDC) by Magnussen \cite{magnussen1981}\cite{magnussen1989}\cite{gran1996}. It is based on the analysis of turbulent scales in the energy cascade and may include arbitrarily complex reaction mechanisms. An extensive review comprising recent developments is available in \cite{boesenhofer2018} and recapitulated synoptically in this section. \newline
In general and as basis for the EDC, the energy cascade hypothesis assumes that mechanical energy is continuously transformed from mean flow to the largest eddies, which subsequently pass it on to increasingly smaller eddies. In addition, it is expected that viscous dissipation mostly takes place within the smallest eddies, which are related to the Kolmogorov scales \cite{pope2000}. These timescales are furthermore connected to chemical reaction processes since they depict perfectly mixed species at a molecular level, a necessary condition for reaction processes. To transfer these presumptions into a numerical model, each computational cell is divided into a fine structure part of Kolmogorov scale size and a secondary segment representing the surrounding fluid. Consequently, each cell comprises a reacting part (fine structures) and a non-reacting one (surrounding fluid). \newline
Typical length and velocity scales are assigned to the fine structure region and are denoted by $L^*$ and $u^*$, which by definition are in the same order of magnitude as the Kolmogorov scales \cite{ertesvag2000}. When setting the characteristic scales of the fine structures in relation to the turbulent flow scales, the fine structure length fraction $\gamma_L$ is defined as
\begin{equation}
	\gamma_L = \frac{u^*}{u'} = \left( \frac{3 C_{D2}}{4 C_{D1}^2} \right)^{1/4} \left( \frac{\nu \epsilon}{k^2} \right) ^{1/4} \approx \frac{L^*}{L'}.
	\label{eq:one}
\end{equation}
$C_{D1}$ and $C_{D2}$ are constants related to strain rate and energy transfer in the cascade model and designated to correctly represent as many flow regimes as possible. For simplification, a directly proportional EDC modeling constant $C_{\gamma}$ as well as the turbulent Reynolds number are introduced:
\begin{equation}
	\gamma_L= C_\gamma \left( \frac{\nu \epsilon}{k^2} \right)^{1/4} = C_\gamma Re_t ^{-1/4}
\end{equation}
Since the fine structures also exchange mass with themselves, the mass fraction occupied by them within each cell was redefined as $\gamma^* = \gamma_L^2$ instead of the natural choice $\gamma^* = \gamma_L^3$ \cite{magnussen1989}. Then, the mass transfer rate between fine structures and their surrounding fluid, in relation to the fine structure mass, is expressed by
\begin{equation}
	\dot{m}^* = 2 \frac{u^*}{L\matchsupheight^*} = \left(\frac{3}{C_{D2}}\right)^{1/2} \left(\frac{\epsilon}{\nu}\right)^{1/2}.
\end{equation}
Therefore, the characteristic time scale of fine structures is denoted by
\begin{equation}
		\tau^* = \frac{1}{\dot{m}^*} = \left( \frac{C_{D2}}{3} \right) ^ 2 \left( \frac{\nu}{\epsilon} \right) ^ {1/2},
\end{equation}
which again may be rewritten in terms of a linear constant $C_{\tau}$ and turbulent Reynolds number
\begin{equation}
	\tau^* = C_{\tau} \left( \frac{\epsilon}{\nu} \right) ^{1/2} = C_{\tau} Re_t^{-1/2} \frac{k}{\epsilon}.
\end{equation}
Default values of the EDC constants equal $C_{\gamma} = 2.1366$ and $C_{\tau} = 0.4082$. By application of the fine structure mass fraction, the mean mass transfer rate in relation to total mass, which can be understood as the mean rate of molecular mixing \cite{ertesvag2000}, is derived as
\begin{equation}
	\dot{m} = \dot{m}^* \gamma^* = \frac{\gamma_L^2}{\tau^*}.
	\label{eq:mdot}
\end{equation}
Mean fluid quantities are expressed by linear combination
\begin{equation}
	\overline{\Psi} = \Psi^0 ( 1 - \gamma_L^3) + \Psi^* \gamma_L^3 .
\end{equation}
Where superscript $0$ denotes properties of the surrounding fluid in each cell. Utilizing this formulation the species' mass fractions are expressed by
\begin{equation}
	Y_i^0 = \frac{\overline{Y_i} - Y_i^* \gamma_L^3}{1 - \gamma_L^3}.
	\label{eq:Yi0}
\end{equation}
When $Y_i^*$ corresponds to species mass fraction in the fine structure region after numerical integration of all chemical reactions, the individual reaction of each specie is expressed as
\begin{equation}
	\overline{\dot{\omega_i}} = \overline{\rho} \, \dot{m} \, \chi (Y_i^* - Y_i^0).
\end{equation}
While $\chi$ denotes the fraction of fine structures that actually participate in the reaction and is usually assumed to unity. Substitution of equations (\ref{eq:mdot}) and (\ref{eq:Yi0}) yields the reaction rate depending on mean quantities as
\begin{equation}
\overline{\dot{\omega_i}} = \overline{\rho} \, \frac{\gamma_L^2}{\tau^* (1 - \gamma_L^3)}  (Y_i^* - \overline{Y_i}).
\end{equation}
The most computationally expensive task of reactive flow simulations is numerical integration of chemical kinetics. To improve this limitation, the reacting fine structure region is solved by a zero-dimensional model. Originally, a perfectly stirred reactor (PSR) was proposed for this purpose \cite{magnussen1981}. It features an isothermal and isobaric set of ODEs, which is solved to steady-state. Due to the persisting extensive numerical cost of this operation, the selected solver follows a different approach by neglecting mixing between fine structures and surrounding to form a simplified set of ODEs referred to as a plug flow reactor (PFR) \cite{boesenhofer2018}. Instead of finding a steady state solution, the PFR ODE system is only integrated for the fine structure time scale $\tau^*$. Boesenhofer \cite{boesenhofer2018} pointed out that there are only minor differences in accuracy. \newline
Two limiting scenarios have to be considered when looking at wall bounded flows in combination with the EDC combustion model. Since turbulent kinetic energy approaches zero at the wall, $\gamma_L$ advances towards infinity which has to be omitted. An elegant solution to eliminate this behavior is to apply blending functions for $\gamma_L$ depending on $\mathrm{Re_t}$ \cite{myhrvold2003}. A more straightforward approach is to apply a limiter directly onto $\gamma_L$, which performs comparably well and is thus employed \cite{myhrvold2003}. \newline 
A second singularity similar to the first one is encountered when $\gamma_L$ approaches unity. This corresponds to solely fine structures in a cell and a species reaction rate approaching infinity. De \cite{de2011} came to the conclusion that the EDC changes the mean species mass fraction by linear relaxation, thus avoiding the complex nonlinear problem. In this context the relation of the linear mixing problem's timescale is described as 
\begin{equation}
\frac{\tau^*}{\tau_{mix}} = \frac{\gamma_L^2}{1 - \gamma_L^3} < 1 .
\end{equation}
Therefore, the stability condition and limiting value for the fine structure mass fraction, which requires the reaction time $\tau^*$ to be smaller than the mixing time scale, equals $\gamma_L = 0.755$. \newline
As the flame-wall interaction (FWI) mode of the configuration is sidewall quenching (SWQ), reactions right at the wall have to be inhibited to properly reflect physical quenching of the flame. Since the EDC model is inhibited in its ability to account for this effect and instead assigns the described limiter, the reaction is prevented by setting the reactions source terms to zero in all wall adjacent cells. Although default EDC modeling constants perform well in a broad application range, they are modified to fit all operating conditions of the featured reactive flow as accurately as possible. An increased constant value of $C_{\gamma} = 3.0$ results in a better agreement to most experimental measurements. The same value is already reported in \cite{de2011}.

\section{Chemical reaction kinetics evaluation}
\begin{figure}[H]
	\centering
	\includegraphics[width=0.68\textwidth]{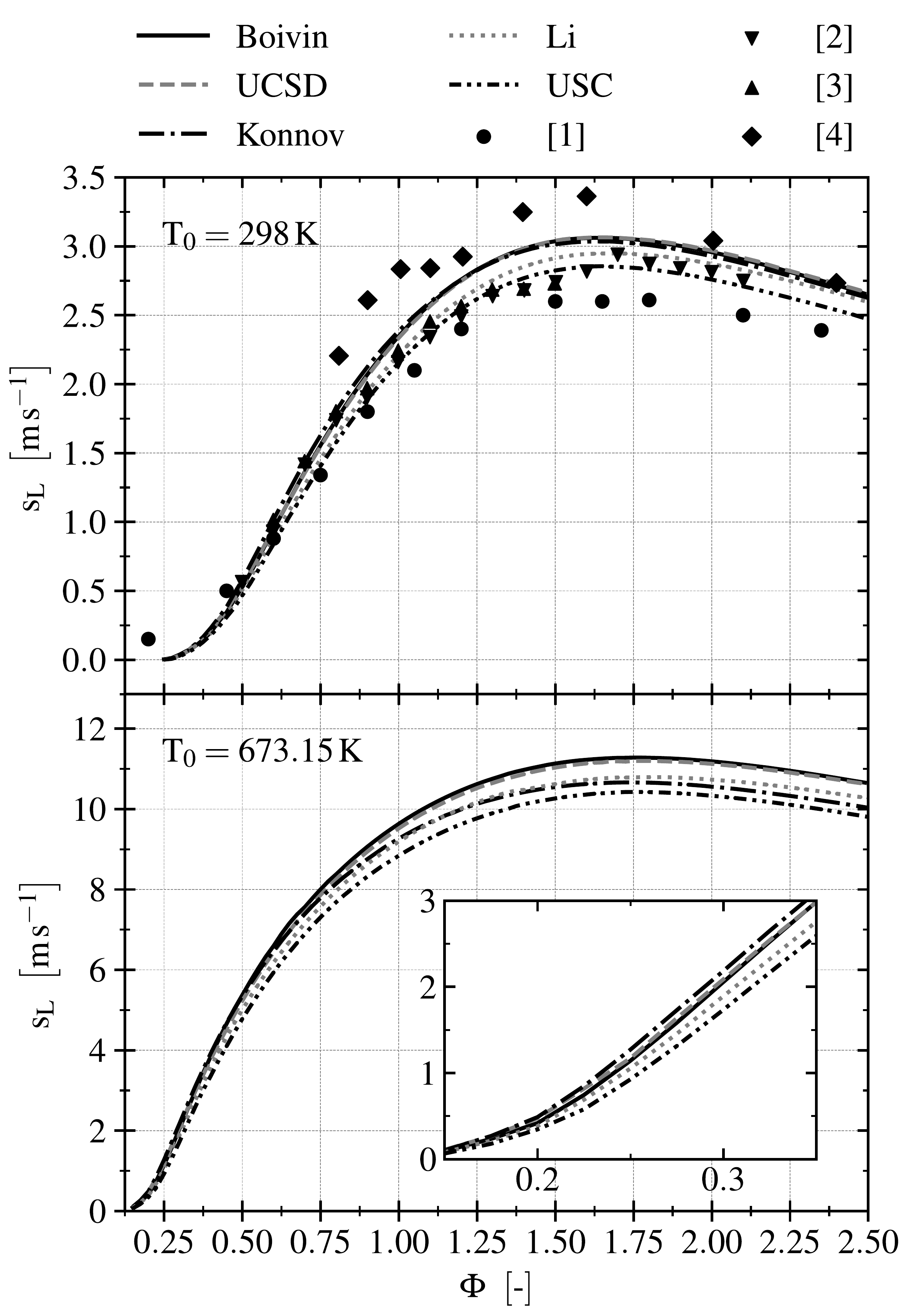}
	\caption{Laminar flame speed results for varying unburnt gas temperatures. Applied linestyles remain valid for all plots in this section.}
	\label{fig:adiabaticPremixedFlame}
\end{figure}

Before conducting a detailed numerical investigation featuring finite rate chemistry, a review of available reaction mechanisms is worthwhile. For that purpose, chemical kinetics are evaluated in a simplified canonical configuration corresponding to a steady one-dimensional premixed flame. To ensure that each mechanism is able to reproduce the central physical phenomenon of the configuration, it has to be part of the canonical problem as well \cite{bioche2019}. 
In the featured setup flame shape, position and extinction behavior are particularly determined by heat transfer processes at the wall. Therefore, the investigation of reaction mechanisms includes non-adiabatic premixed flame calculations under heat loss conditions. Although hydrogen/air kinetics have been explored for an number of different applications \cite{liu2019} \cite{vincentRandonnier2019} \cite{kumar2015}, this study distinctly focuses on the influence of heat loss and derived quenching behavior. \newline
For comparison five reaction mechanisms are taken into account. Four of them are considered detailed (Konnov \cite{konnov2015}, Li \cite{li2004}, UCSD \cite{ucsd}, USC \cite{davis2005}) while the final one is reduced (Boivin \cite{boivin2011}). They collectively feature four identical major species (\chemfig{H_2}, \chemfig{O_2}, \chemfig{N_2}, \chemfig{H_2O}) as well as five minor ones (\chemfig{OH}, \chemfig{H_2O_2}, \chemfig{H}, \chemfig{O}, \chemfig{HO_2}). The number of total reactions varies from $12$ to $29$. Due to their much larger chemical time scale, nitrogen reactions are neglected in all calculations. A more detailed perspective on the different kinetics is summarized in table \ref{tab:chemKinetic}. One of the key differences is that the simplified mechanism comprises irreversible reactions while the detailed ones refrain from it. \newline
The experimental test setup allows for \chemfig{OH^{*}} chemiluminescence measurements. That is why in addition to hydrogen oxidation, an \chemfig{OH^{*}} chemiluminescence sub-mechanism has to be included into the overall reaction kinetic. For the study at hand, the sub-mechanism of Kathrotia \cite{kathrotia2010} is incorporated. 
\begin{table}[H]
	\caption{Distribution of reaction types of investigated hydrogen/air reaction mechanisms.}
	\centering
	\def\arraystretch{1.3}
	\begin{spacing}{1.1}\centering
		\begin{tabular}{@{} l c c c c @{}}
			\toprule
			& $\mathrm{N_{Total}}$ & $\mathrm{N_{Fall\mhyphen off}}$ & $\mathrm{N_{Pressure}}$ & $\mathrm{N_{Irreversible}}$ \\ 		
			\midrule
			Konnov \cite{konnov2015} & 29 & 4 & 4 & 0 \\
			Li \cite{li2004}         & 21 & 2 & 4 & 0 \\
			UCSD \cite{ucsd}         & 23 & 2 & 4 & 0 \\
			USC \cite{davis2005}     & 25 & 2 & 4 & 0 \\
			Boivin \cite{boivin2011} & 12 & 2 & 2 & 6 \\
			\bottomrule
		\end{tabular}
	\end{spacing}
	\label{tab:chemKinetic}
\end{table}
The simplified numerical formulations are solved by the Cantera laminar reacting flow solver \cite{cantera}. The initial problem of reduced complexity is an adiabatic one-dimensional freely propagating flame. This configuration is well suited to examine laminar flame speed as an inherent property of reaction kinetics. The influence of equivalence ratio on laminar flame velocity is depicted and compared to its experimental counterpart in figure \ref{fig:adiabaticPremixedFlame}. At standard conditions the deviation between experiment and numerical calculations for all featured mechanisms, lies within the usually reported error margin. It should be noted that all simplified calculations are carried out under atmospheric pressure conditions since the impact of pressure variations in the burner setup are considered subordinate. The series is repeated for elevated unburnt mixture temperatures to investigate settings closer to the actual operating conditions. As detailed in figure \ref{fig:adiabaticPremixedFlame} there are only minor differences for lean flame fronts ranging from $\mathrm{\Phi=0.2}$ to $\mathrm{\Phi=0.3}$. Over the entire equivalence ratio range the largest disparity is observed between the detailed Konnov and reduced Boivin chemical kinetics. 
Since wall heat transfer and subsequent quenching reflect the major physical phenomenon, a canonical configuration incorporating these effects is analyzed. For this purpose the energy equation of the laminar flame solver is altered to incorporate an additional heat sink term. In fact, the heat sink is realized by introducing a heat loss factor $\mathrm{\gamma_{loss}}$ to the already implemented heat of reaction source term. Therefore, the complete energy equation in mixture fraction space is expressed by:
\begin{equation}
\overline{\rho}\: \overline{c_p} u \frac{d T}{dZ} = \frac{d}{dZ} (\overline{\lambda} \frac{d T}{d Z}) - \sum_{k}^{n_k}j_k c_{p,k} \frac{d T}{dZ} - (1-\gamma_{loss}) \sum_{k}^{n_k} h_k W_k \dot{\omega}_k
\end{equation}
By inclusion of this heat loss factor only the reaction front of the premixed flame is impacted while the remaining flame remains unaltered \cite{proch2015}. \newline
\begin{figure}[H]
	\centering
	\includegraphics[width=0.72\textwidth]{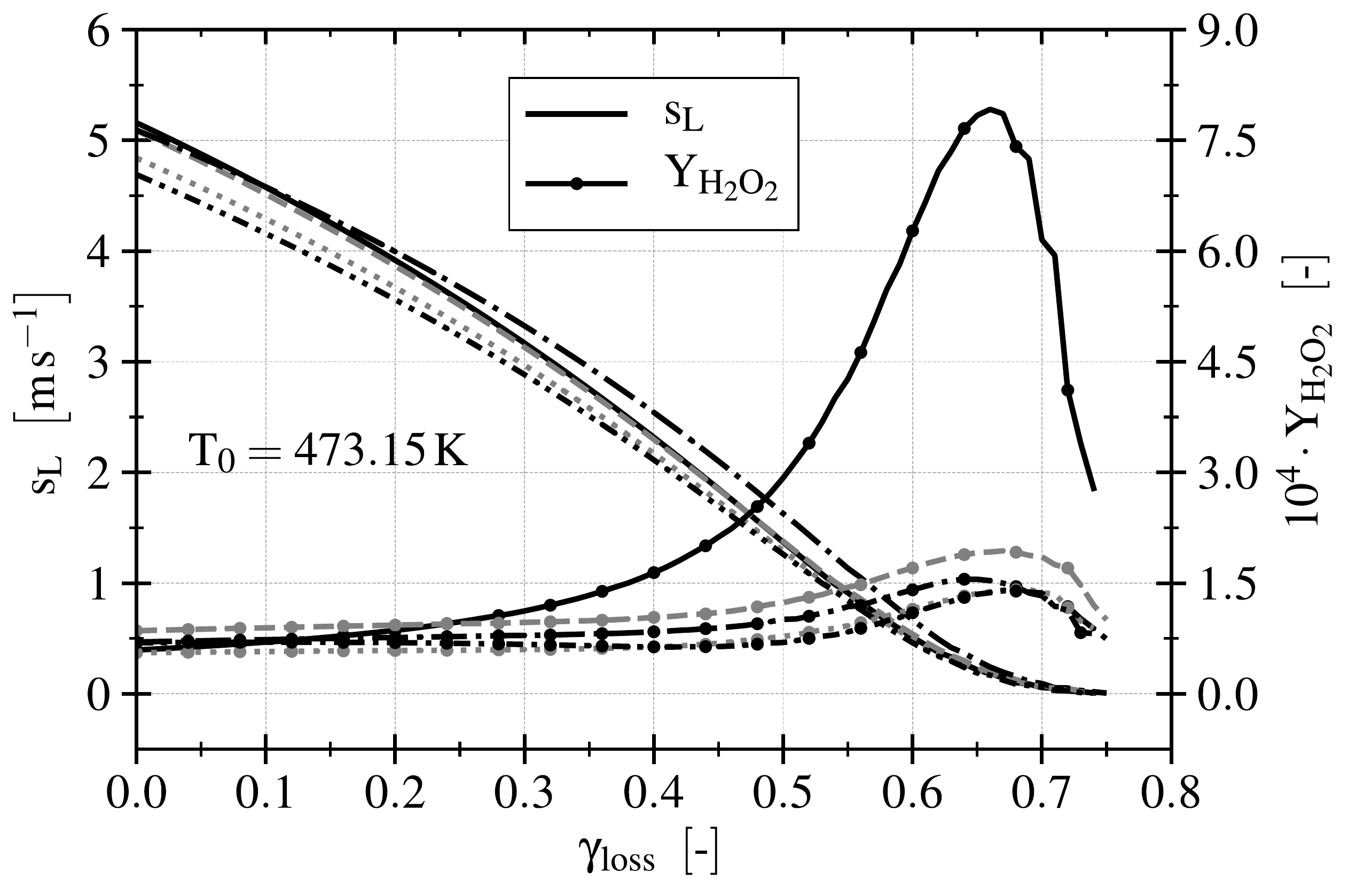}
	\caption{Laminar flame speed under the influence of external heat loss.}
	\label{fig:sl_by_gamma_loss}
\end{figure}

\begin{figure}[H]
	\centering
	\hspace*{-3.5cm}
	\includegraphics[width=1.6\textwidth]{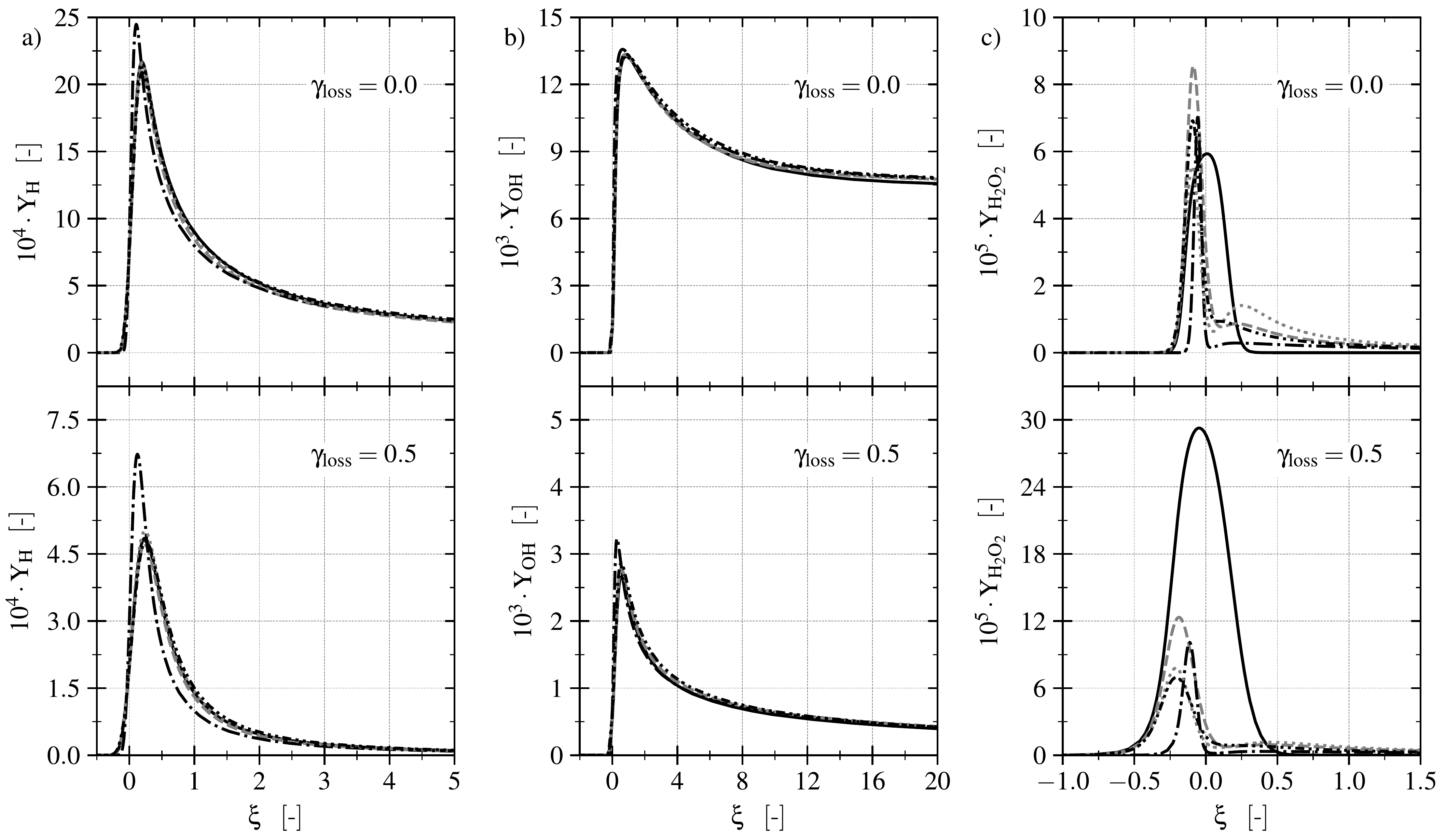}
	\caption{Minor species' mass fractions under adiabatic and non-adiabatic conditions $\left(\mathrm{\gamma_{loss}} = 0.5\right)$.}
	\label{fig:heatLossPremixedFlame}
\end{figure}

Figure \ref{fig:sl_by_gamma_loss} shows the influence of the established heat loss factor on laminar flame velocity as well as hydrogen peroxide concentration as an intermediate species. In general, quenching is observed at a heat loss factor of $\mathrm{\gamma_{loss} = 0.745}$. Discrepancies in laminar flame speed between the applied chemical kinetics become smaller as the extinction point is approached. Contradictorily, deviations in \chemfig{H_2O_2} mass fraction between detailed and  simplified mechanisms intensify for greater heat loss factors. \newline
This effect is further investigated in figure \ref{fig:heatLossPremixedFlame} with help of a non-dimensional variable to simplify visualization of the narrow reaction zone:
\begin{equation}
	\xi = \frac{x - x_{ff}}{\Delta_{ff}}
\end{equation}
Where $x_{ff}$ and $\Delta_{ff}$ denote the flame front's positions as well as thickness. The figure visualizes minor species \chemfig{H}, \chemfig{OH} and \chemfig{H_2O_2} under adiabatic and non-adiabatic imposed heat loss conditions. Species plots 5a and 5b show a clear trend of diminished intermediate species concentration under heat loss conditions while all employed reaction kinetics behave fairly similar. This again changes for the hydrogen peroxide concentration. Firstly, additional heat loss does not seem to significantly alter the maximum species concentration when any of the detailed kinetics is employed (fig. 5c). A second observation is that the ma	ximum \chemfig{H_2O_2} mass fraction under heat loss conditions is about four times higher for the simplified mechanism compared to detailed ones. \newline
Although the sceletal mechanism of Boivin performs comparably well in adiabatic calculations, it demonstrated discrepancies for minor species under heat loss conditions. Therefore, the more computationally expensive reactions mechanism by Li is selected for the numerical study. It should be noted that the other detailed chemical kinetics perform very similar to the Li mechanism while including a higher number of reactions. Therefore, in addition to the chemiluminescence reactions, ten species and $29$ reactions are considered in all consecutive simulations. 

\section{Numerical investigation}
\label{sec:numericalStudyAndResults}
An examination of numerical aspects of wall anchoring turbulent flames in the featured configuration is presented in the following chapter. It focuses on the influence of mesh resolution, choice of chemical kinetics as well as RANS-based turbulence modeling. Furthermore, a comprehensive review of the general flame structure and flame/wall interaction by SWQ is carried out. 

\subsection{Flame structure and wall anchoring phenomenon}
\begin{figure}[H]
	\centering
	\hspace*{-1.2cm}
	\includegraphics[width=1.2\textwidth]{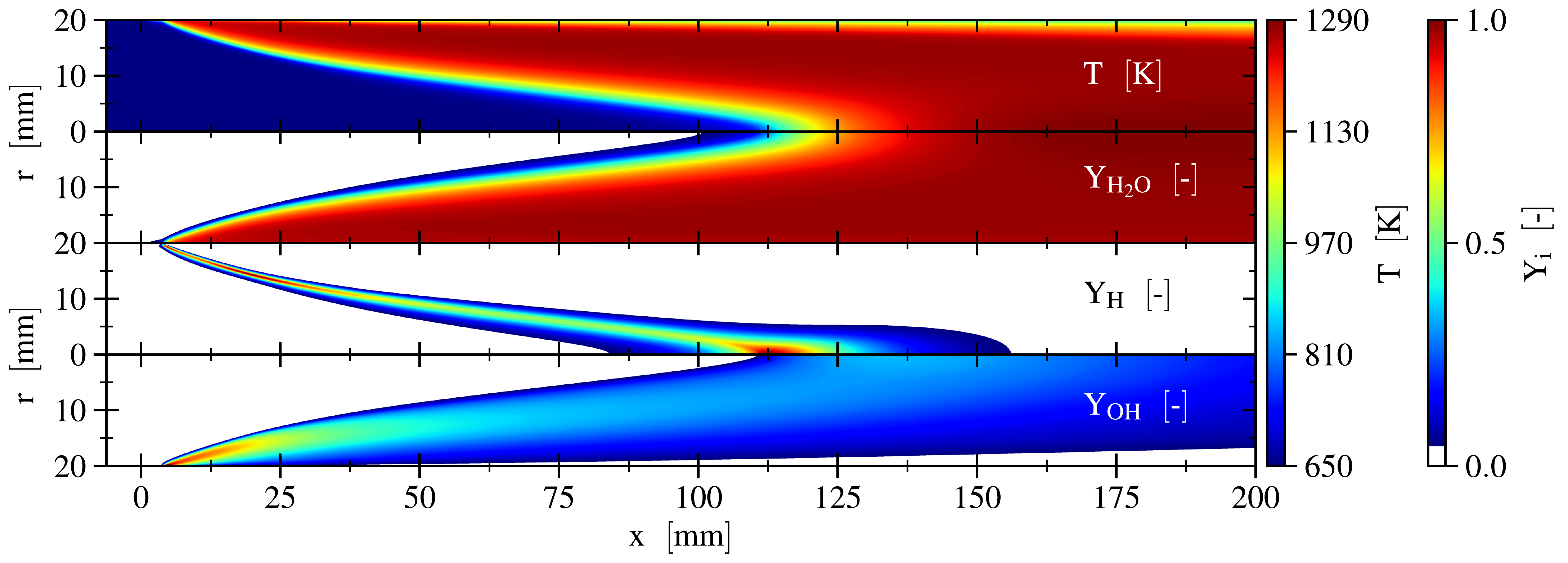}
	\caption{General flame structure for the operating conditions of case 1.}
	\label{fig:generalFlameStructure}
\end{figure}

\begin{figure}[H]
	\centering
	\includegraphics[width=0.8\textwidth]{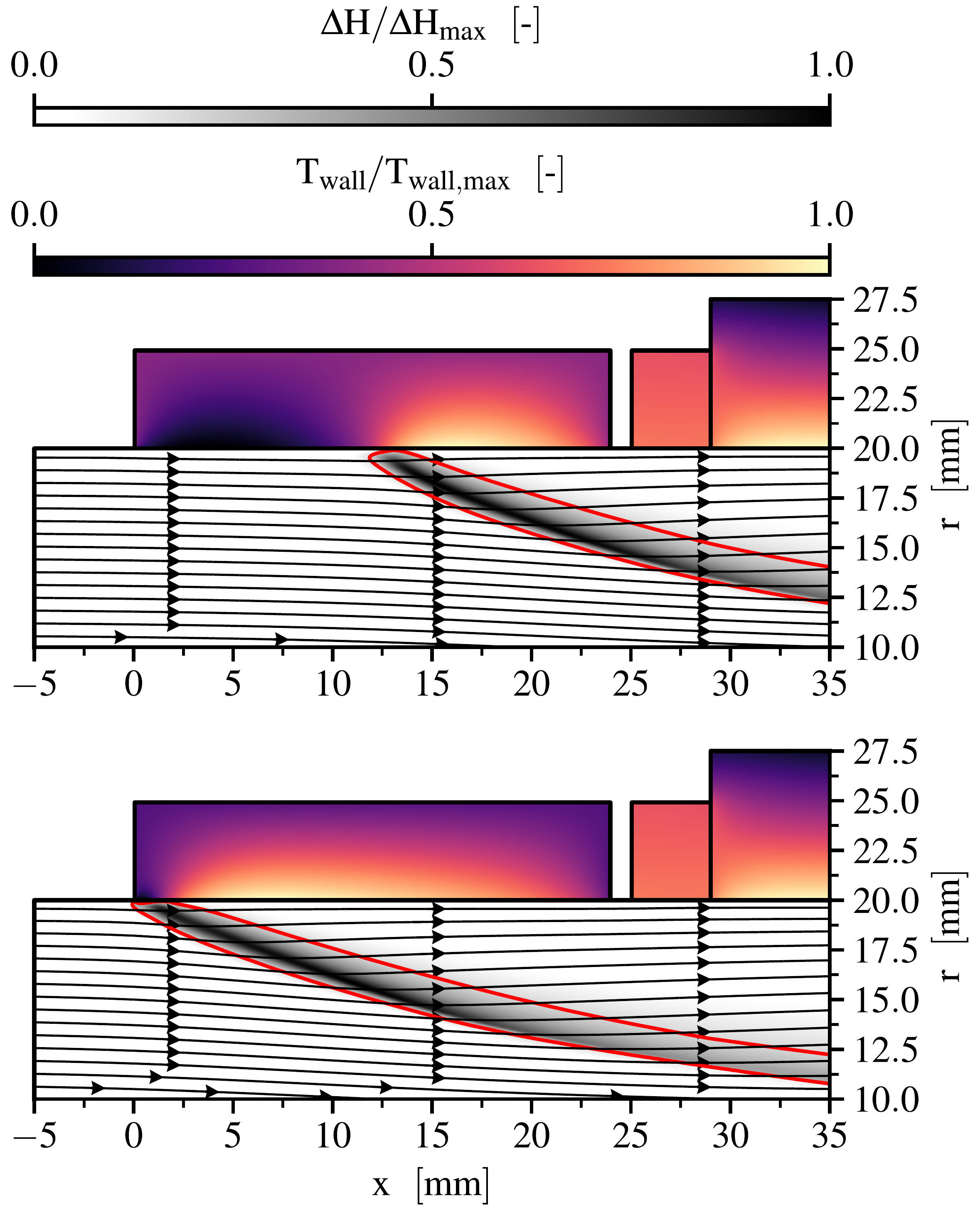}
	\caption{Flame wall interaction for a hot wall temperature of $\mathrm{T_{w,hot} = \SI{720}{\celsius}}$ (top) and $\mathrm{T_{w,hot} = \SI{870}{\celsius}}$ (bottom). Otherwise, applied inlet boundary conditions reflect case 1. Red isolines depict $\mathrm{\SI{5}{\percent}}$ of the maximum heat release rate.}
	\label{fig:detailsAtWall}
\end{figure}

The general flame shape and composition are displayed in figure \ref{fig:generalFlameStructure} by depicting flame temperature and various species distributions. To demonstrate their differences, the major species $\chemfig{H_2O}$, minor species $\chemfig{OH}$ as well as the quickly recombining species $\chemfig{H}$ are portrayed. The featured boundary conditions correspond to the previously defined case 1, which is utilized as a reference case for most of the numerical investigation in this chapter. As the portrayed coordinate system corresponds to the one defined in figure \ref{fig:grid}, the flame anchors steadily shortly after the front face of the externally heated wall segment. \newline
As briefly discussed in the introduction, steady flame/wall anchoring is introduced by interaction of reaction products and fresh reactants with adjacent walls. In detail, hot reaction products transfer heat into neighboring walls downstream of the flame front by convection and radiation. In return, heat conduction within the solid is induced. This mechanism is responsible for increasing the wall's temperature upstream of the flame front, which eventually leads to heat transfer into the fluid right in front of the flame front. Therefore, ignition of the unburnt gas is induced more effortlessly as well as continuously. \newline
The described phenomenon by itself may already lead to steady flame anchoring. Additional external heating of a secluded wall segment enhances the anchoring process and forces it to take place at a specific wall location. In the described numerical setup this behavior is imposed by application of constant temperature boundary conditions at all heated wall segments (fig. \ref{fig:grid}). Thus, heat introduced into the modeled heating ring may only escape back into the fluid or locally increase the walls temperature. \newline
Figure \ref{fig:detailsAtWall} displays this behavior by visualizing wall temperature, heat of reaction as well as velocity streamlines. Due to disparate temperature scales between externally heated and non-heated walls, each wall's temperature distribution is normalized separately. Red isolines depict $\mathrm{\SI{5}{\percent}}$ of the maximum heat release rate and therefore correspond to the beginning and end of the reaction zone. All featured operating conditions (tab. \ref{tab:operatingConditions}) feature measured hot wall temperatures of $\mathrm{\SI{800}{\kelvin}}$. As the moderate flame anchoring position of case 1 is very similar in most considered operating conditions, two more extreme hot wall conditions are investigated. These correspond to an imprinted hot wall temperature of $\mathrm{\SI{720}{\kelvin}}$ (fig. \ref{fig:detailsAtWall} top) and $\mathrm{\SI{870}{\kelvin}}$ (fig. \ref{fig:detailsAtWall} bottom). \newline
From the displayed results it may be inferred that the unburnt mixture is ignited immediately at the beginning of the heated segment if excessive wall temperatures are present. Under these conditions additional heat conduction through the wall, induced by the hot flame products, seems to be redundant. Nonetheless, elevated wall temperatures still enhance the reaction's flammability limits and lead to more stable flame anchoring. As expected, the contrary behavior is observed for a decrease in heated wall temperature. In that case, a stable flame anchoring position is achieved further downstream. Although the influence of external heating means on the flame anchoring position is apparent, it has no significant impact on flame shape as well as general flame composition. A general requirement for the presented flame anchoring technique is that turbulent flame speed as well as flow velocity match reasonably well.

\subsection{Influence of mesh resolution and chemical kinetics}
\begin{figure}[H]
	\centering
	\includegraphics[width=1.0\textwidth]{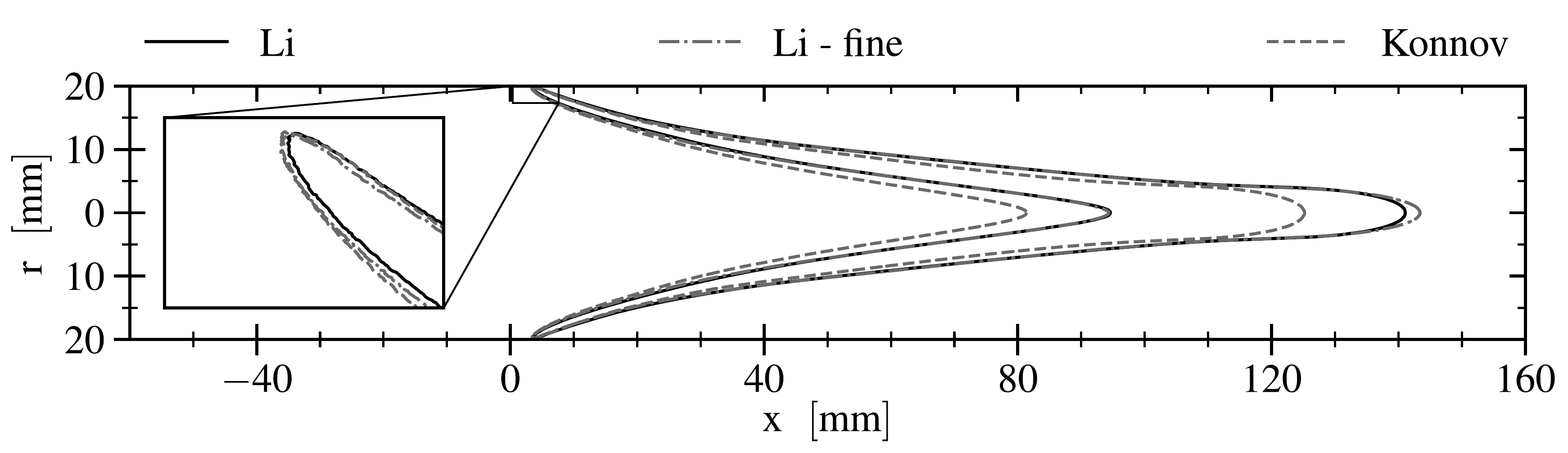}
	\caption{Influence of grid resolution and kinetic mechanism on reaction front's shape. Applied inlet boundary conditions reflect case 1. Isolines depict  $\mathrm{\SI{5}{\percent}}$ of the reaction's maximum heat production.}
	\label{fig:gridResolutionAndKinetics}
\end{figure}

To verify the chosen numerical setup, a mesh study featuring a refined grid is conducted. For the fine mesh, the resolution is doubled compared to the original grid. This leads to a minimum cell size at the wall of $\mathrm{\SI{20}{\micro\meter}}$ and a total cell count of $\mathrm{5.96\cdot10^6}$ quadrangular cells. As depicted in figure \ref{fig:gridResolutionAndKinetics} the finer mesh does not lead to significant deviations in flame shape or anchoring position. Thus, the utilized mesh may be considered adequate to reflect all relevant physical phenomena. \newline 
Since the preceding examination of chemical kinetics is performed for one-dimensional canonical problems, the impact of different mechanisms on the more realistic two dimensional flame shape and anchoring position is reviewed. For that purpose the selected mechanism by Li is compared to the more complex kinetics by Konnov. Figure \ref{fig:gridResolutionAndKinetics} shows a slightly shortened flame front for the Konnov mechanism while the anchoring position is not affected significantly. Therefore, the smaller flame speed already observed in figure \ref{fig:adiabaticPremixedFlame}, is also discernible in two-dimensional axisymmetric simulations. But, as both mechanisms lie within the experimentally observed flame speeds, the much more efficient mechanism by Li is still considered suitable. 

\subsection{RANS-based turbulence modeling}
To solve the closure problem of time-averaged Navier-Stokes equations, multiple RANS turbulence models are applied to operational conditions resembling case 1. A detailed performance evaluation is given in the following. \newline
Two-equation turbulence models have already been successfully applied in RANS simulations of premixed combustion \cite{wan2015}. This work aims to extend the assessment of turbulence models in cooperation with the EDC for wall anchoring flames under SWQ conditions. Due to their inherent ability to accurately resolve turbulent boundary layers, $\mathrm{\omega\mkern 1mu \mhyphen}$based models are exclusively adopted for the study at hand. For this purpose the standard, baseline (BSL) and shear stress transport (SST) $\mathrm{k\mhyphen\omega}$ models are selected as classical two-equation models. In addition, an $\mathrm{\omega\mkern 1mu \mhyphen}$based Reynolds stress turbulence model featuring five transport equations is evaluated as well. In conclusion, the investigated RANS turbulence models are: $\mathrm{k\mhyphen\omega \: std}$, $\mathrm{k\mhyphen\omega \: BSL}$, $\mathrm{k\mhyphen\omega \:SST}$ and stress-$\mathrm{\omega}$ RSM. For all $\mathrm{\omega\mkern 1mu \mhyphen}$based turbulence models, which can be integrated through the entire viscous sub-layer, wall treatment is identical as well as $\mathrm{y^+}$-insensitive. Therefore, and since the applied mesh satisfies the $\mathrm{y^+ \leq 1}$ condition for all operating conditions, no turbulent wall functions are applied at any point. \newline
\begin{figure}[H]
	\centering
	\hspace*{-1.2cm}
	\includegraphics[width=1.2\textwidth]{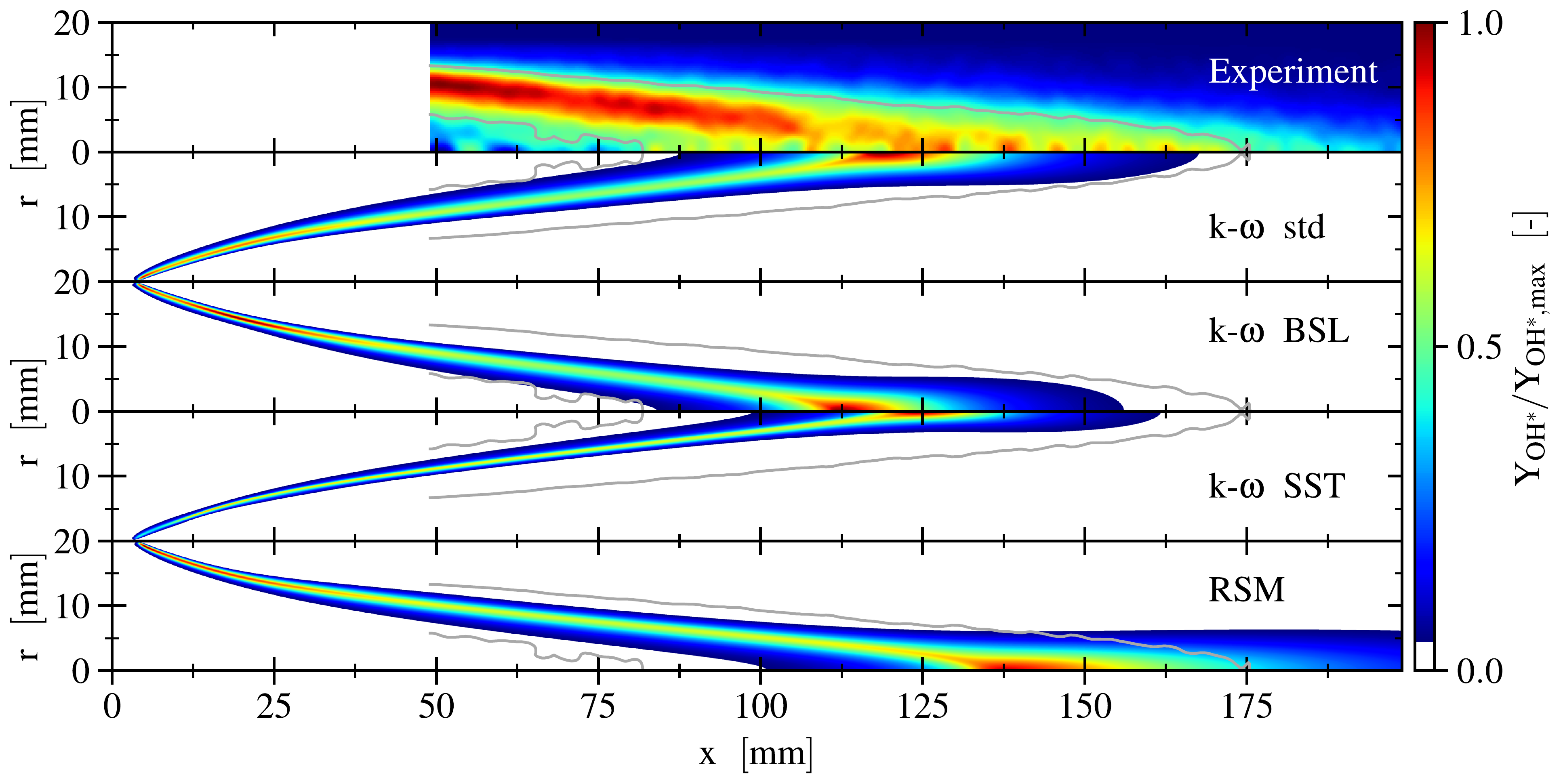}
	\caption{Influence of RANS-based turbulence models. Setup corresponds to case 1. Experimental results are visualized by inverse Abel transformation. Grey lines correspond to iso-values of $0.5$.}
	\label{fig:turbulenceModels}
\end{figure}
A qualitative comparison of Abel inversed experimental measurements and simulation results featuring the different turbulence models is shown in figure \ref{fig:turbulenceModels}. Satisfactory overall agreement is achieved by all employed turbulence models. One minor difference concerns varying flame anchoring positions. This is visualized in detail in figure \ref{fig:surfaceHeatFlux} by depiction of total surface heat flux over the heated wall segment. It shows that there is only a minor discrepancy between  $\mathrm{k\mhyphen\omega \:std}$, $\mathrm{k\mhyphen\omega \:BSL}$ and RSM model while the $\mathrm{k\mhyphen\omega \:SST}$ model differs more significantly. \newline
Another more pronounced dissimilarity concerns the location of the maximum concentration of the \chemfig{OH^{*}} chemiluminescence radical. Except for the $\mathrm{k\mhyphen\omega \:SST}$ model, the employed turbulence models predict the maximum \chemfig{OH^{*}} concentration to be at the flame anchoring position. This is in accordance with the inverse Abel transformed experimental data. In contrast, the $\mathrm{k\mhyphen\omega \:SST}$ model computes the maximum location to be at the flame's tip and thus symmetry line. Furthermore, the later model predicts different maximum values of intermediate species. This behavior is visualized by the distribution of the \chemfig{H} mass fraction along the axis of symmetry (fig. \ref{fig:H_at_symmetry}). \newline
Due to its elongated flame tip, it could be argued that the RSM model, which promises to be the most accurate turbulence model directly calculating the Reynolds stress tensor's components, reflects the experimental measurements most accurately. But due to its extensive computational cost, the $\mathrm{k\mhyphen\omega \:BSL}$ model is selected for all subsequent simulations during the experimental validation for different operating conditions. It should be noted that the $\mathrm{k\mhyphen\omega \:std}$ compares equally well. 

\begin{figure}[H]
	\centering
	\includegraphics[width=0.68\textwidth]{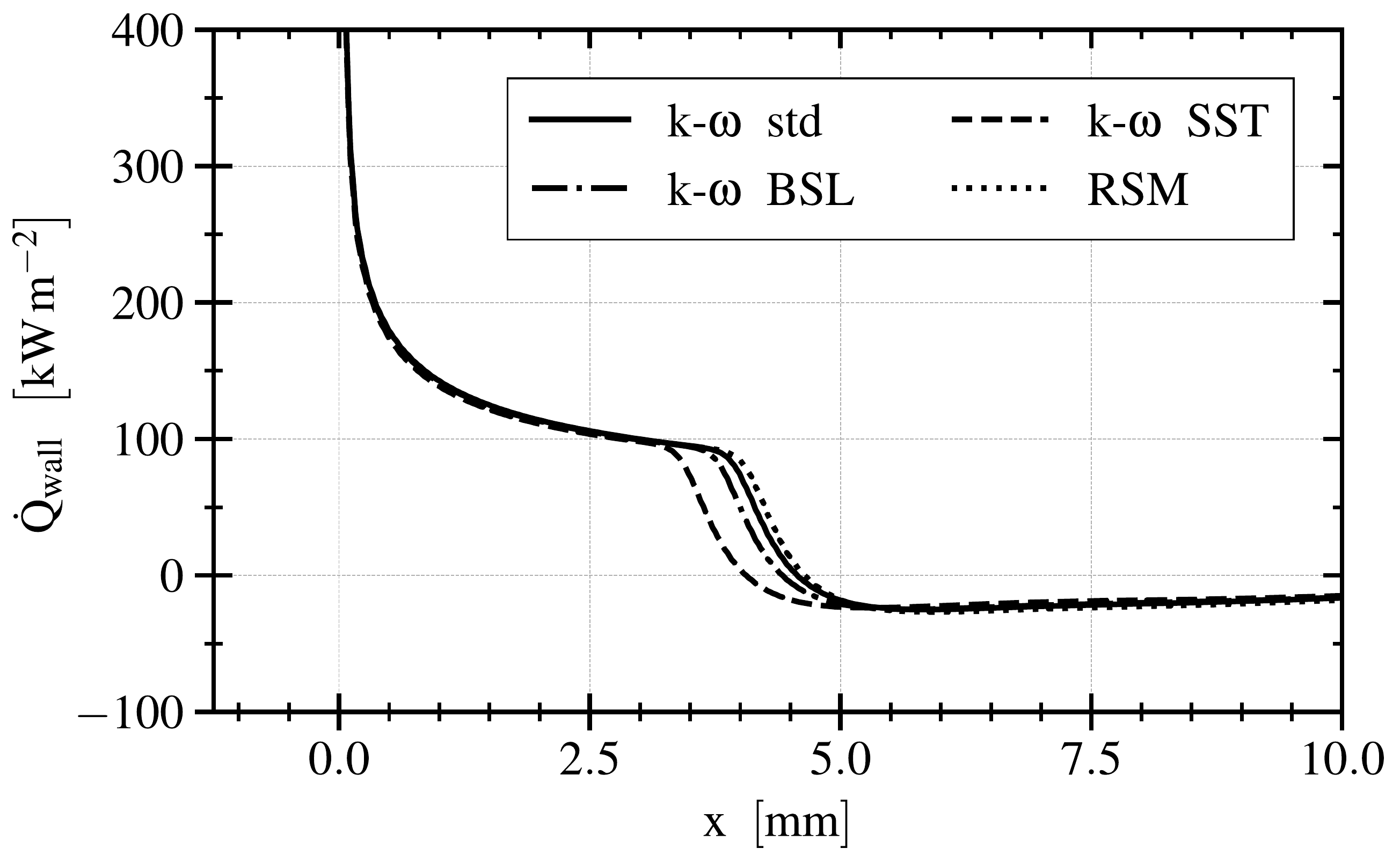}
	\caption{Surface heat flux on externally heated wall segment for different turbulence models.}
	\label{fig:surfaceHeatFlux}
\end{figure}
\begin{figure}[H]
	\centering
	\includegraphics[width=0.68\textwidth]{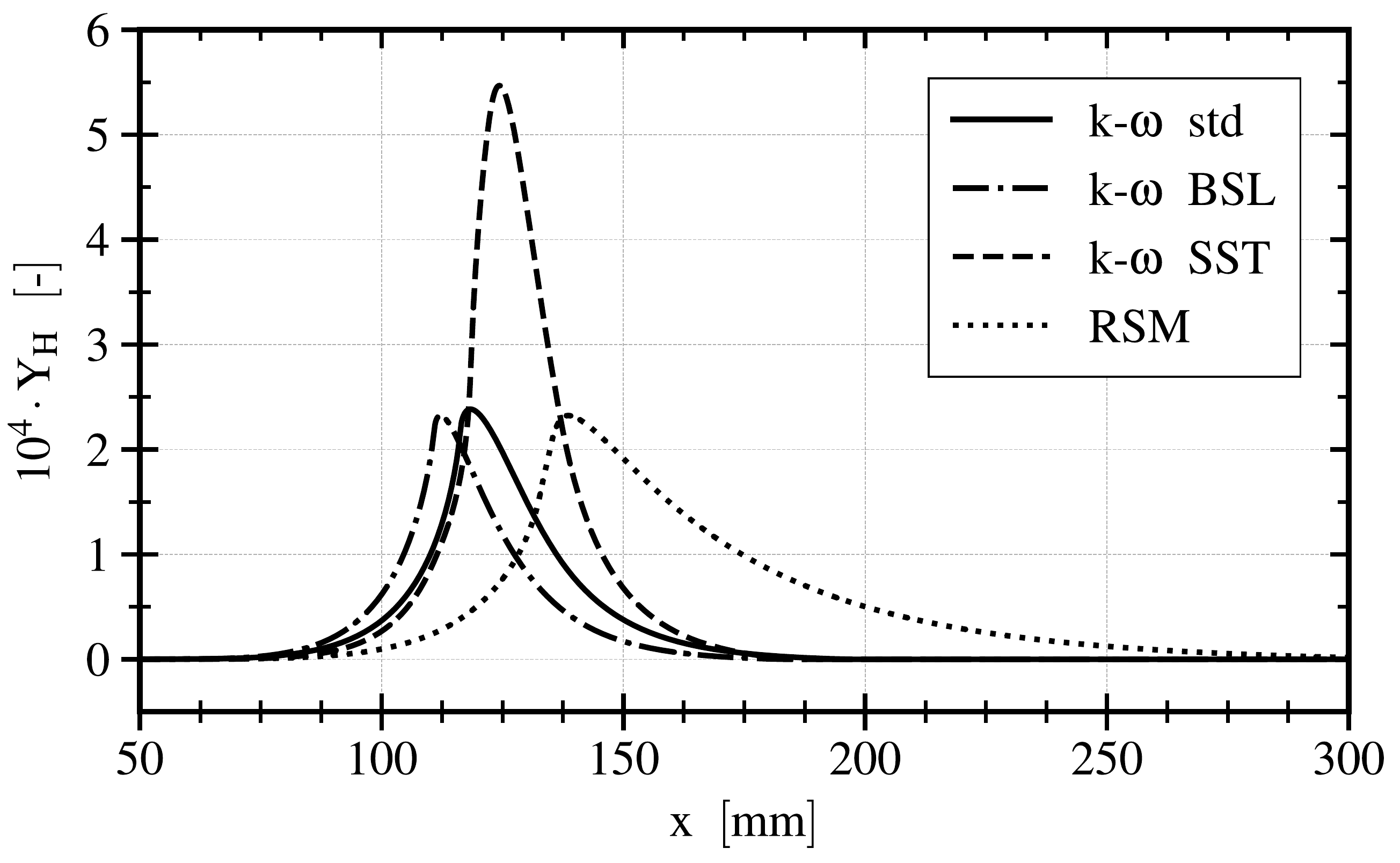}
	\caption{Distribution of hydrogen atom concentration along the flame's symmetry line.}
	\label{fig:H_at_symmetry}
\end{figure}

\section{Experimental validation}
\begin{figure}[H]
	\centering
	\hspace*{-1.2cm}
	\includegraphics[width=1.2\textwidth]{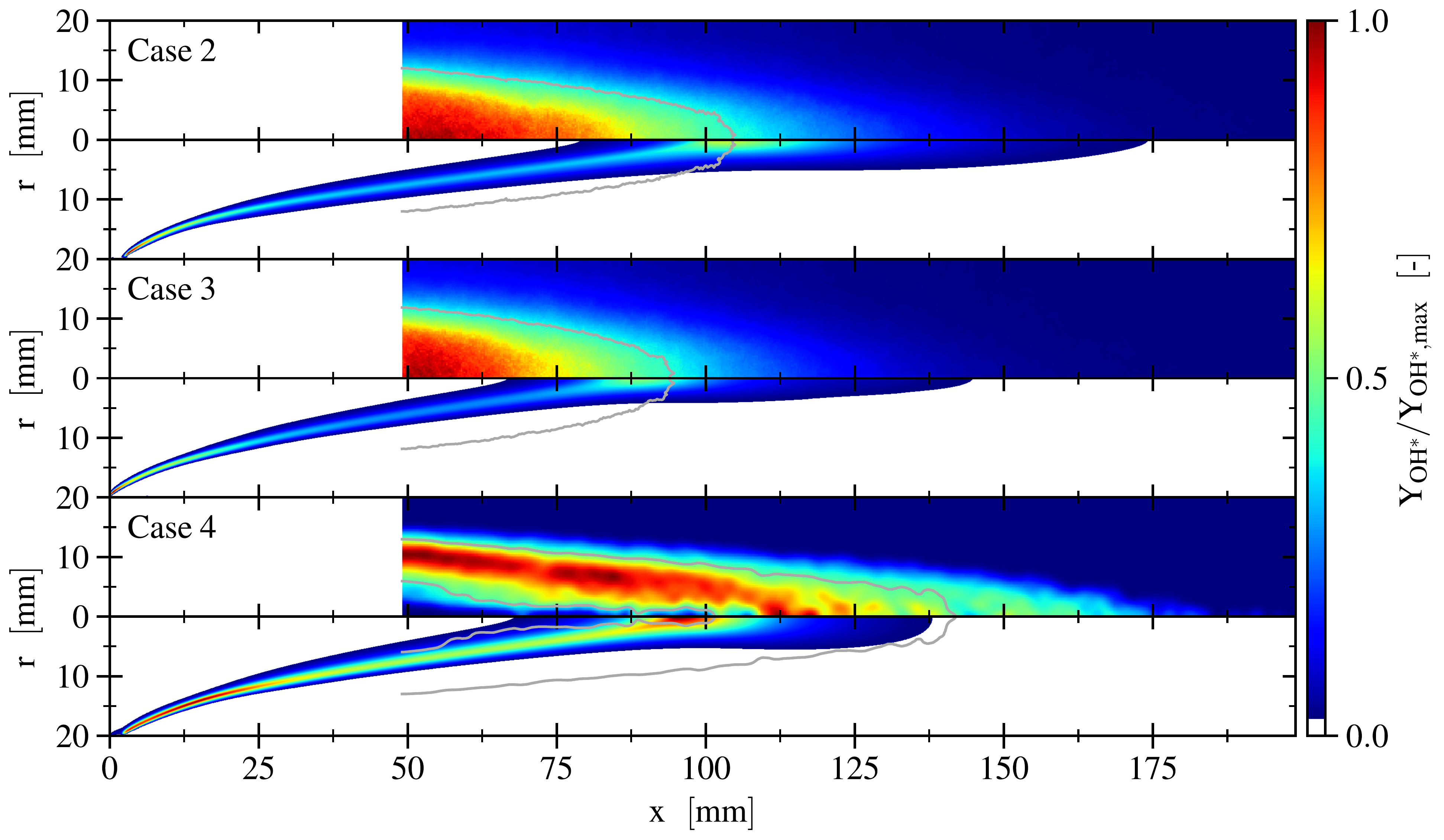}
	\caption{Influence of equivalence ratio and bulk Reynolds number visualized by simulation as well as chemiluminescence measurements.}
	\label{fig:equivalenceRatioAndReynoldsNumber}
\end{figure}

\begin{figure}[H]
	\centering
	\hspace*{-1.2cm}
	\includegraphics[width=1.2\textwidth]{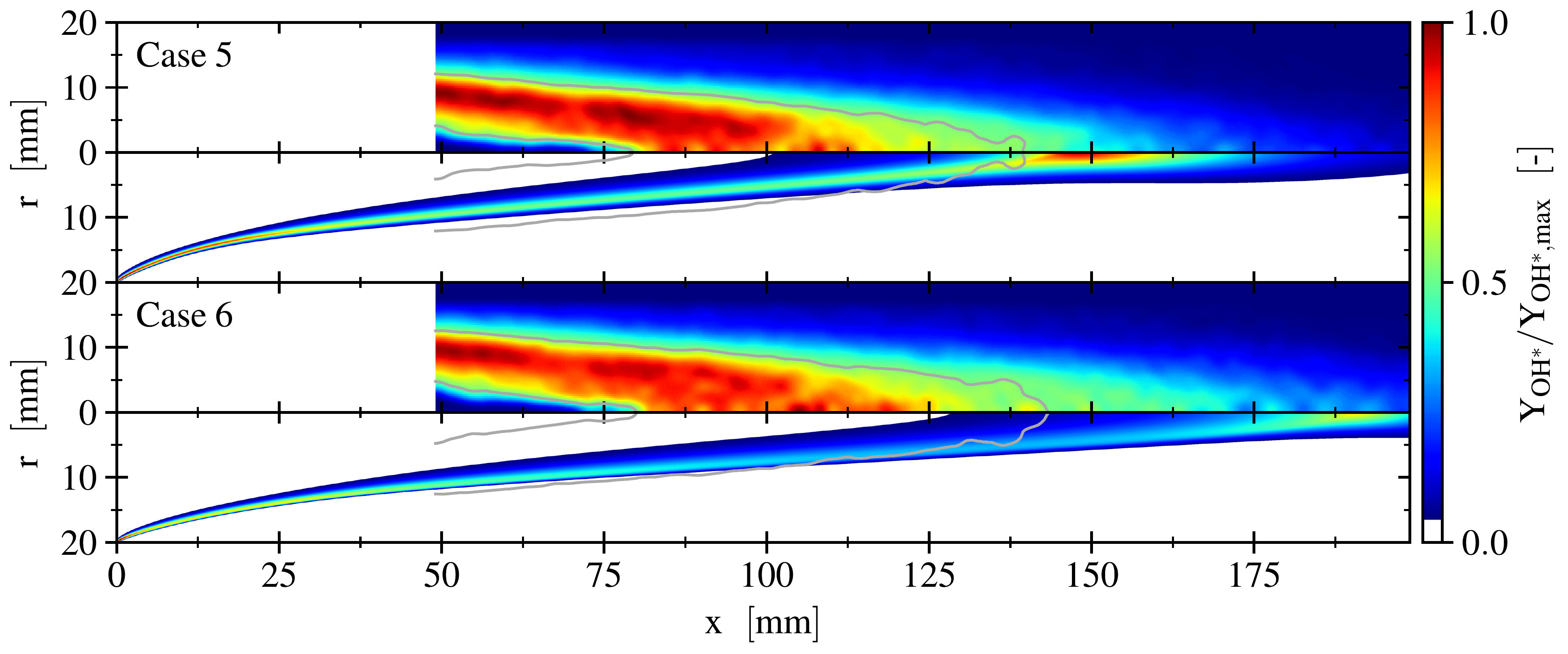}
	\caption{Influence of reduced unburnt mixture temperature visualized by simulation as well as Abel inversed chemiluminescence measurements.}
	\label{fig:inletTemperature}
\end{figure}

Besides the already assessed operating conditions of case 1, the numerical setup is further validated by rigorous comparison of numerical results and experimental measurements for various operating conditions. Again, chemiluminescence of the \chemfig{OH^{*}} radical is utilized to bring measurements and calculations together. Featured operating points are summarized in table \ref{tab:operatingConditions}. These are chosen to investigate the influence of three main quantities: equivalence ratio, bulk Reynolds number and unburnt mixture temperature. A special focus is put on the universal applicability of the EDC modeling constants. \newline 
Results showing the impact of differences in equivalence ration as well as in bulk Reynolds number are visualized in figure \ref{fig:equivalenceRatioAndReynoldsNumber}. By increasing the inlet hydrogen mass fraction, the equivalence ratio is adjusted from $\Phi = 0.2$ (case 1) to $\Phi=0.3$ (case 2). This induces an enhanced flame speed which results in a shortened flame length. This effect is visualized for an inlet mass flow rate of $\mathrm{80\;m/s}$ (case 2) and $\mathrm{60\;m/s}$ (case 3 and case 4). The flame shortening is clearly pronounced for both cases. Qualitative agreement between experiments and numerical results is apparent for all cases in figure \ref{fig:equivalenceRatioAndReynoldsNumber} while the EDC modeling constants remain unaltered ($C_{\gamma} = 3.0$ and $C_{\tau} = 0.4082$). Only minor deviations in flame length and reaction zone location are evident. \newline 
Due to inevitable constraints in the experimental setup, the measured section for some inlet conditions only depicts the flame's tip, thus rendering an inverse Abel transformation impossible. Therefore, numerical results are compared to symmetrically averaged but otherwise unaltered chemiluminescence measurements. This is visualized in the high equivalence ratio cases in figure \ref{fig:equivalenceRatioAndReynoldsNumber}. Thereby, the corresponding isolines solely depict the end of the reaction zone. \newline
Finally, results for operating conditions reflecting case 4 and case 5 are discussed. These illustrate the influence of colder temperatures of the unburnt gas. Such an inflow enthalpy reduction of the mixture culmintaes in much lower turbulent flame speeds and thus elongated flames. This behavior is clearly shown in both the experiment as well as numerical simulations (fig. \ref{fig:inletTemperature}). Compared to the previously discussed operating conditions, the simulations predict a longer flame front than observed in corresponding experiments. This could be remedied by slight modification of EDC modeling constants. \newline

\section{Conclusions}
A thermal flame holder design suitable for continuous combustion devices has been studied experimentally as well as numerically. Its application to premixed hydrogen/air flames was evaluated in detail by consideration of a diverse set of operating conditions. Prior to experimental validation of the numerical setup, the influence of different kinetic mechanisms, mesh resolution and RANS-based turbulence models was assessed. The study of chemical kinetics under heat loss conditions concluded that there are only minor discrepancies between the featured detailed mechanisms while a reduced one expressed serious deviations while experiencing heat loss. \newline
An experimental validation by qualitative \chemfig{OH^{*}} chemiluminescence comparison was carried out. For that purpose optical \chemfig{OH^{*}} measurements are either compared directly or after an inverse Abel transformation to the numerical chemiluminescence distribution. The validation shows that the applied numerical method is able to predict the general flame shape and anchoring behavior quite well. It was worked out that an adaptation of the inherent EDC combustion modeling constant $C_{\gamma}$ to $C_{\gamma} = 3.0$ provides superior agreement. This applies to various inlet mass flow rates their respective bulk Reynolds numbers and imprinted equivalence ratios. Only for extensively elongated flames by reduction of the unburnt mixture temperature, deviations between experiment and numerical results become more pronounced.

\section{Acknowledgements}
The authors gratefully acknowledge the Gauss Centre for Supercomputing e.V. (www.gauss-centre.eu) for funding this project by providing computing time on the GCS Supercomputer SuperMUC at Leibniz Supercomputing Centre (www.lrz.de). Furthermore, the research project is part of the SPP 1980 founded by the German Research Foundation (DFG), whose support the authors thankfully acknowledge. 

\section*{References}

\bibliography{mybibfile}

\end{document}